
\input epsf.tex

\font\rmu=cmr10 scaled\magstephalf
\font\bfu=cmbx10 scaled\magstephalf

\font\it=cmti10 scaled \magstephalf

\rmu

\font\rmus=cmr8
\font\rmuss=cmr6
\font\mait=cmmi10 scaled\magstephalf
\font\maits=cmmi7 scaled\magstephalf
\font\maitss=cmmi7
\font\msyb=cmsy10 scaled\magstephalf
\font\msybs=cmsy8 scaled\magstephalf
\font\msybss=cmsy7
\font\bfus=cmbx7 scaled\magstephalf
\font\bfuss=cmbx7
\font\cmeq=cmex10 scaled\magstephalf

\textfont0=\rmu
\scriptfont0=\rmus
\scriptscriptfont0=\rmuss

\textfont1=\mait
\scriptfont1=\maits
\scriptscriptfont1=\maitss

\textfont2=\msyb
\scriptfont2=\msybs
\scriptscriptfont2=\msybss

\textfont3=\cmeq
\scriptfont3=\cmeq
\scriptscriptfont3=\cmeq

\newfam\bmufam  \textfont\bmufam=\bfu
      \scriptfont\bmufam=\bfus \scriptscriptfont\bmufam=\bfuss

\hsize=15.5cm
\vsize=22cm
\baselineskip=16pt   
\parskip=12pt plus  2pt minus 2pt

\def\a{\alpha}
\def\b{\beta}
\def\d{\delta}

\def\g{\gamma}

\def\semi{\bigcirc\kern-1em{s}\;}

\def\del{\partial}
\def\ni{\noindent}
\def\R{{\rm I\!R}}

\def\one{{\mathchoice {\rm 1\mskip-4mu l} {\rm 1\mskip-4mu l}
{\rm 1\mskip-4.5mu l} {\rm 1\mskip-5mu l}}}
\def\Q{{\mathchoice
{\setbox0=\hbox{$\displaystyle\rm Q$}\hbox{\raise 0.15\ht0\hbox to0pt
{\kern0.4\wd0\vrule height0.8\ht0\hss}\box0}}
{\setbox0=\hbox{$\textstyle\rm Q$}\hbox{\raise 0.15\ht0\hbox to0pt
{\kern0.4\wd0\vrule height0.8\ht0\hss}\box0}}
{\setbox0=\hbox{$\scriptstyle\rm Q$}\hbox{\raise 0.15\ht0\hbox to0pt
{\kern0.4\wd0\vrule height0.7\ht0\hss}\box0}}
{\setbox0=\hbox{$\scriptscriptstyle\rm Q$}\hbox{\raise 0.15\ht0\hbox to0pt
{\kern0.4\wd0\vrule height0.7\ht0\hss}\box0}}}}
\def\C{{\mathchoice
{\setbox0=\hbox{$\displaystyle\rm C$}\hbox{\hbox to0pt
{\kern0.4\wd0\vrule height0.9\ht0\hss}\box0}}
{\setbox0=\hbox{$\textstyle\rm C$}\hbox{\hbox to0pt
{\kern0.4\wd0\vrule height0.9\ht0\hss}\box0}}
{\setbox0=\hbox{$\scriptstyle\rm C$}\hbox{\hbox to0pt
{\kern0.4\wd0\vrule height0.9\ht0\hss}\box0}}
{\setbox0=\hbox{$\scriptscriptstyle\rm C$}\hbox{\hbox to0pt
{\kern0.4\wd0\vrule height0.9\ht0\hss}\box0}}}}

\font\fivesans=cmss10 at 4.61pt
\font\sevensans=cmss10 at 6.81pt
\font\tensans=cmss10
\newfam\sansfam
\textfont\sansfam=\tensans\scriptfont\sansfam=\sevensans\scriptscriptfont
\sansfam=\fivesans
\def\sans{\fam\sansfam\tensans}
\def\Z{{\mathchoice
{\hbox{$\sans\textstyle Z\kern-0.4em Z$}}
{\hbox{$\sans\textstyle Z\kern-0.4em Z$}}
{\hbox{$\sans\scriptstyle Z\kern-0.3em Z$}}
{\hbox{$\sans\scriptscriptstyle Z\kern-0.2em Z$}}}}

\newcount\foot
\foot=1
\def\note#1{\footnote{${}^{\number\foot}$}{\ftn #1}\advance\foot by 1}

\def\frac#1#2{{#1\over #2}}
\def\text#1{\quad{\hbox{#1}}\quad}

\font\ch=cmbx12 scaled\magstephalf
\font\ftn=cmr8 scaled\magstephalf

\font\it=cmti10 scaled\magstephalf

\font\titch=cmbx12 scaled\magstep2
\font\titname=cmr10 scaled\magstep2
\font\titit=cmti10 scaled\magstep1
\font\titbf=cmbx10 scaled\magstep2

\nopagenumbers


\line{\hfil }
\line{\hfil CGPG-94/7-1}
\line{\hfil July 20, 1994}
\vskip3cm
\centerline{\titch INDEPENDENT LOOP INVARIANTS FOR}
\vskip.5cm
\centerline{\titch 2+1 GRAVITY}
\vskip2cm
\centerline{\titname R. Loll\note{address after Aug 94: Sezione INFN di
Firenze, Largo E. Fermi 2, Arcetri, I-50125 Firenze, Italy}}
\vskip.5cm
\centerline{\titit Center for Gravitational Physics and Geometry}
\vskip.2cm
\centerline{\titit Physics Department, Pennsylvania State University}
\vskip.2cm
\centerline{\titit University Park, PA 16802, U.S.A.}

\vskip4cm
\centerline{\titbf Abstract}
We identify an explicit set of complete and independent Wilson loop invariants
for 2+1 gravity on a three-manifold $M=\R\times\Sigma^g$, with $\Sigma^g$ a
compact oriented Riemann surface of arbitrary genus $g$. In the derivation
we make use of a global cross section of the $PSU(1,1)$-principal bundle over
Teichm\"uller space given in terms of Fenchel-Nielsen coordinates.

\vfill\eject
\footline={\hss\tenrm\folio\hss}
\pageno=1


\line{\ch 1 Introduction\hfil}

In recent years, Wilson loop variables have found manifold application in
the investigation of fundamental physical theories. This holds in particular
for the canonical quantization of gravity, and for quantum chromodynamics,
the latter both in the continuum and the regularized lattice version (see
[1] for a review of loop methods). Still many aspects of these loop
formulations remain poorly understood. On the one hand they enjoy explicit
gauge invariance, but on the other a vast overcompleteness is introduced by
considering Wilson loops corresponding to arbitrary closed curves in
space(-time). This obscures the physical content of the theory, which becomes
particularly obvious in explicit calculations as for instance those occurring
in lattice gauge theory. There the overcompleteness of the Wilson loop
variables constitutes the biggest obstacle to making efficient numerical
computations [2].

There are in general topological obstructions to finding sets of loop
variables which are both independent and complete on the quotient space
${\cal A}/\cal G$ of connection one-forms modulo gauge transformations.
Although $\cal A$ possesses an affine, i.e. an ``almost linear" structure,
this is in general not any more true for the corresponding space of $\cal
G$-orbits in $\cal A$. This is the case for both $SU(N)$-gauge theory and
for 3+1 gravity in the Ashtekar formulation, where the gauge group is
$SL(2,\C)$.

One notable
exception to this state of affairs is the theory of 2+1 gravity. This theory
has been studied both for its mathematical beauty and as a toy model for the
(3+1)-dimensional theory, in particular, to gain more understanding of
quantum loop representations for gravitational theories [3,4]. In the pure
(2+1)-dimensional theory, without matter coupling, the absence of local
field degrees of freedom may be ``compensated" for by allowing the underlying
space-time manifold $M$ to have a non-trivial topology. We will be concerned
with the case $M=\R\times\Sigma^g$, where $\Sigma^g$ is a compact orientable
Riemann surface of genus $g\geq 2$. Using a space of connections as the basic
variables of the theory,  it could be demonstrated that its physical (i.e.
classically reduced) configuration space is both finite-dimensional and
contractible, and isomorphic to the $(6g-6)$-dimensional Teichm\"uller space
[5]. Thus it is possible in principle to find a set of loop variables that
can serve as globally well-defined coordinates on the reduced configuration
space.

Although there have been extensive investigations of both the classical and
quantum theory of the genus-1 case (which from a mathematical point of view
is somewhat degenerate), not a great deal is known about the explicit
physical dynamics for higher genus. From the point of view of the loop
quantization approach, it is important to understand which features of the
genus-1 case generalize to higher $g$. One of the first steps in achieving
this is to understand how Wilson loop variables describe the physical
configuration (and phase) space. This problem can be phrased as follows.
The reduced configuration space may be described as (a sector of) the space
${\cal A}^F$  of flat $SO(2,1)$-connections  $A(x)$ on $\Sigma^g$ modulo the
group $\cal G$ of $SO(2,1)$-gauge transformations. Because of their gauge
invariance, the Wilson loop variables

$$
T(\g):= {\rm Tr}\; U_\g ={\rm Tr \;P}\exp \oint_\g A \eqno(1.1)
$$

\ni are functions on the quotient space ${\cal A}^F/\cal G$. Since the
connections are flat, the Wilson loop (1.1) depends only on the homotopy
equivalence class of the closed curve $\g$. However, Wilson loop variables
corresponding to arbitrary elements $\g$ of the homotopy group $\pi_1
(\Sigma^g)$ are not all independent but subject to i) identities among the
traces of products of $SO(2,1)$-representation matrices, so-called
Mandelstam constraints, and ii) identities coming from the fundamental
relation

$$
\a_1\b_1\a_1^{-1}\b_1^{-1}\a_2\b_2\a_2^{-1}\b_2^{-1}\dots \a_g\b_g\a_g^{-1}
\b_g^{-1}=1 \eqno(1.2)
$$

\ni between the generators $\a_i$, $\b_i$, $i=1\dots g$ of the homotopy
group (s. Fig.1). (Recall that a general homotopy group element $\g$ is a
``word" in terms of the $\a_i$ and $\b_i$, i.e. a finite ordered product of
the generators and their inverses.)  The difficulty now lies in identifying
a relevant set  of such identities and solving them to obtain $6g-6$
independent trace invariants. For $g=2$, an explicit solution has been given
by Nelson and Regge [6]. The computations are rather involved and their
result so far has not been extended to genus $g>2$.

\vskip1cm
\epsffile{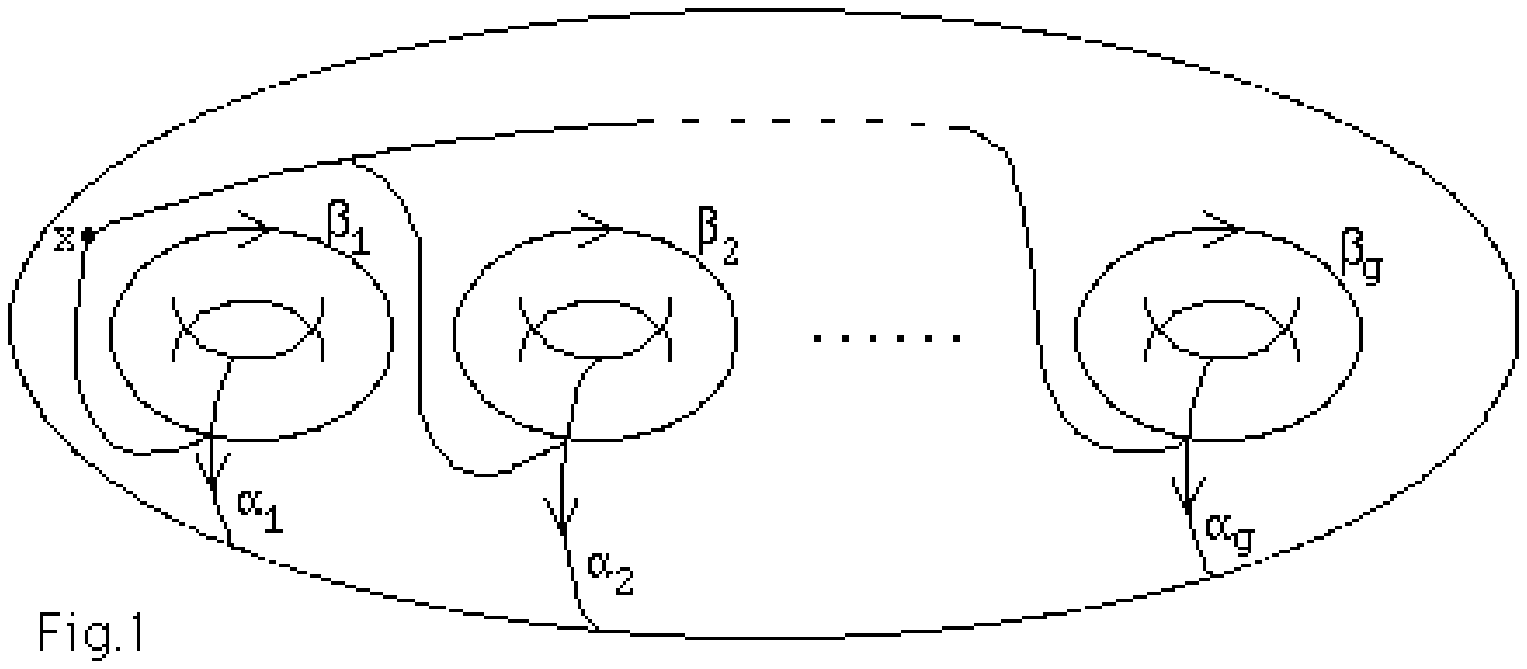}

In this paper, we will present a solution for the general genus case.
However, instead of tackling the algebra of the Wilson loop constraints
directly, we will make use of the explicit parametrization of the
Teichm\"uller spaces in terms of Fenchel-Nielsen coordinates and of a result
by Okai, namely, an explicit global section of the principal
$PSU(1,1)$-bundle over Teichm\"uller space [7]. This enables us to express
arbitrary Wilson loops as functions of the Fenchel-Nielsen coordinates. It
then remains to find an algebraically independent set of such Wilson loops.
We give one solution, i.e. $6g-6$ independent (linear combinations of) Wilson
loops for any genus $g$, which have a particularly simple algebraic
dependence on the Fenchel-Nielsen coordinates.

\vskip 1.5cm

\line{\ch 2 The loop invariants\hfil}

We will now briefly review the necessary ingredients for deriving our result.
The Riemann-Hilbert action for three-dimensional Lorentzian gravity on a
space-time manifold $M=\R\times\Sigma^g$ may be substituted by an action on
a space of connections, which requires the introduction of internal
$SO(2,1)$-degrees of freedom. After a Legendre transformation one obtains a
first-class constrained system, which remarkably can be solved explicitly (for
more details on the general theory, see [5,8]). As already mentioned in the
introduction, the  reduced configuration space may be described as a quotient
space ${\cal A}^F/\cal G$ (and the corresponding physical phase space is the
cotangent bundle over ${\cal A}^F/\cal G$). Alternatively, one may consider
the set of holonomies $U_{\a_i}$, $U_{\b_i}$ around the $2g$ generators of
$\pi_1(\Sigma^g)$ modulo gauge transformations at the common base point
$x\in\Sigma^g$ and subject to

$$
U_{\a_1}U_{\b_1}U_{\a_1}^{-1}U_{\b_1}^{-1}\dots
U_{\a_g}U_{\b_g}U_{\a_g}^{-1}U_{\b_g}^{-1}=\one \eqno(2.1)
$$

\ni which is a direct consequence of relation (1.2). Furthermore, the
holonomies must all lie in the sector of $SO(2,1)$ consisting of boosts
around spacelike axes. For computational simplification we will work in the
two-dimensional representation of $SU(1,1)$, identifying opposite points.
The gauge group is therefore to be identified with $PSU(1,1)=SU(1,1)/\Z_2$,
where we have divided out the normal subgroup. This form enables us to apply
directly a result obtained by Okai [7], who constructed a global section of
the trivial principal bundle

$$
\eqalign{
{\rm Hom}(\pi_1(\Sigma^g&),PSU(1,1))^{e=2g-2}\cr
&\downarrow \cr
{\rm Hom}(\pi_1(\Sigma^g),PSU &(1,1))^{e=2g-2}/PSU(1,1)={\cal T}_g.}
\eqno(2.2)
$$

\ni The space of homomorphisms ${\rm Hom}(\pi_1(\Sigma^g),PSU(1,1))^{e=2g-2}$
is the same as the space of the holonomies $U$ described above, before
factoring out by the gauge group action. The superscript $e=2g-2$ denotes the
connected component consisting of representations whose associated $\R
P^1$-bundle over $\Sigma^g$ has Euler number $2g-2$ [9]. This condition selects
exactly the sector of holonomies we are interested in.
The bundle fibre is given by the group $PSU(1,1)$, acting adjointly at the
base point $x$ of the homotopy generators. The base space of the bundle is
naturally isomorphic to the Teichm\"uller space ${\cal T}_g$ of $\Sigma^g$.

The Teichm\"uller space ${\cal T}_g$ is diffeomorphic to $\R^{6g-6}$ and may
be parametrized globally by the so-called Fenchel-Nielsen coordinates [10].
These are a set of length and angle coordinates associated with a pants
decomposition of the genus-$g$ surface. The surface is cut along $3g-3$ simple
geodesic curves (geodesic with respect to a constant negative-curvature metric
of value $-1$) into $2g-2$ ``pants" $P_i$. As indicated in Figs.1 and 2, the
geodesics will be labelled by $\a_i$, $i=1\dots g$, $\g_i$, $i=1\dots g-1$,
and $\d_i$, $i=2\dots  g-1$. In terms of the homotopy group generators, one
has $\g_i=\b_i \a_i^{-1} \b_i^{-1} \a_{i+1}$ and $\d_i=\g_i\dots \g_{g-1} \b_g
\a_g^{-1}\b_g^{-1}$. With each of these geodesics, we associate a pair of
numbers $(l_j,\tau_j)\in \R^+\times\R$ which measure the length of the geodesic
and the relative twist with which the two pants meeting along the geodesic cut
may be glued together. Following Okai's notation, we associate the index $j$
with the $3g-3$ geodesics as follows: $\a_1$: $j=-\infty$; $\g_1$: $j=0$;
$\d_i$: $j=3i-5$, $\a_i$: $j=3i-4$, $\g_i$: $j=3i-3$ ($2\leq i\leq g-1$);
$\a_g$: $j=\infty$.

\vskip1cm
\epsffile{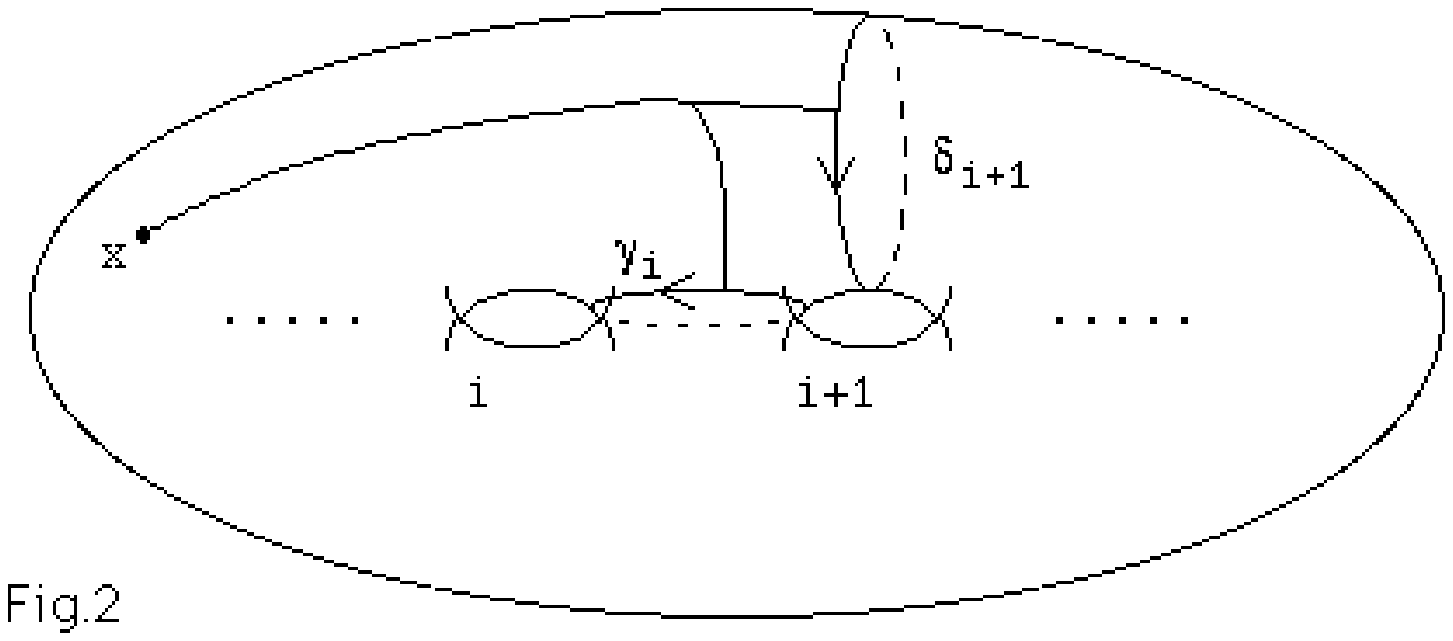}

One proceeds by subdividing each $P_i$ along three geodesic arcs (connecting
pairs of its boundary components) into two right-angled hexagons. The lengths
of these geodesic arcs are not independent but depend on the $l_i$ through
identities coming from hyperbolic geometry. One then associates
$PSU(1,1)$-matrices  to both the geodesic arcs and the boundary components,
depending on their geodesic lengths. To determine the $PSU(1,1)$-holonomy
matrix associated with a given element $\g$ of $\pi_1(\Sigma^g)$, one chooses
a representative that is homotopically  equivalent to $\g$ and is made up
exclusively of such geodesic lines.  Starting at the base point $x$, one
multiplies together the corresponding $PSU(1,1)$-matrices. Crossing over from
one pair of pants to a neighbouring one contributes a $PSU(1,1)$-matrix
depending
on a $\tau_i$. This is described in detail in reference [7], where it is also
proven that this leads to a global cross section of the bundle (2.2). Taking
traces, one obtains arbitrary Wilson loop variables (1.1) as functions of the
$(l_i,\tau_i)$.

Our task is thus reduced to finding an independent, but complete set of
trace invariants that can serve as parameters on ${\cal T}_g$. Obviously it
does not suffice to consider just the Wilson loops of the homotopy generators,
or those along the simple geodesic curves, because this does not lead to the
desired number of $6g-6$ degrees of freedom. One also has to take care that
once a set of ``basic loops" has been found, the corresponding Wilson loops
are indeed good {\it global} coordinates on Teichm\"uller space. For example,
the six invariants for $g=2$ given in [4], based on a set of ``simple-looking"
loops, do not have this property, although they are locally independent.

Since the traces of holonomies essentially measure the lengths of closed
geodesics, it is not hard to extract information about the $3g-3$ length
parameters $l_j$. One just takes the trace of the holonomy along the geodesic
cut $\a_i$, $\g_i$ or $\d_i$. For example, for $\a_1$ one obtains

$$
T(\a_1) = 2 \cosh \frac{l_{-\infty}}{2},\eqno(2.3)
$$

\ni and similarly for the Wilson loops of the remaining geodesic cuts.
The dependence on the twist variables $\tau_j$ is more difficult to extract.
One possible solution will be given below after (2.6). However,
for the sake of illustration, we will first give the set of complete and
independent Wilson loop invariants for $g=2$ and $3$, before writing down the
expressions for the general genus case. Using the abbreviation

$$
s(l_i,l_j,l_k):= \cosh^2\frac{l_i}{2}+\cosh^2\frac{l_j}{2}+\cosh^2\frac{
l_k}{2} +2\,\cosh\frac{l_i}{2}\,\cosh\frac{l_j}{2}\,\cosh\frac{l_k}{2} -1
\eqno(2.4)
$$

\ni for the strictly positive function depending on the length parameters
alone, our solution for $g=2$ is

$$
\eqalign{
&T(\a_1)=2 \cosh \frac{l_{-\infty}}{2}\cr
&T(\b_1\a_1^{-1}\b_1^{-1}\a_2)=2 \cosh\frac{l_0}{2}\cr
&T(\a_2)=2 \cosh \frac{l_{\infty}}{2}\cr
&T(\b_2\a_2\b_2^{-1}\b_1^{-1}\a_2^{-1}\b_1\a_1)-T(\b_2\a_2\b_2^{-1}\a_1
\b_1^{-1}\a_2^{-1}\b_1) =4 \frac{1}{ \sinh\frac{l_{-\infty}}{2} }
s(l_{-\infty},l_0,l_\infty)\,\sinh\tau_{-\infty}\cr
&T(\a_1^{-1}\b_1\a_1^{-1}\b_1^{-1}\a_2\b_1\a_1\b_1^{-1})-T(\a_1^{-1}\a_2)=
4 \frac{1}{ \sinh\frac{l_0}{2} }
s(l_{-\infty},l_0,l_\infty)\,\sinh\tau_0\cr
&T(\b_1\a_1\b_1^{-1}\a_2^{-1}\b_2^{-1}\a_1^{-1}\b_2)- T(\b_1\a_1\b_1^{-1}
\b_2^{-1}\a_1^{-1}\b_2\a_2^{-1}) =
4 \frac{1}{ \sinh\frac{l_\infty}{2} }
s(l_{-\infty},l_0,l_\infty)\,\sinh\tau_\infty,}\eqno(2.5)
$$

\ni and the one for $g=3$

$$
\eqalign{
&T(\a_1)=2 \cosh \frac{l_{-\infty}}{2}\cr
&T(\b_1\a_1^{-1}\b_1^{-1}\a_2)=2 \cosh\frac{l_0}{2}\cr
&T(\a_1\b_1\a_1^{-1}\b_1^{-1}\a_2) =2 \cosh\frac{l_1}{2}\cr
&T(\a_2)=2 \cosh \frac{l_2}{2}\cr
&T(\b_2\a_2^{-1}\b_2^{-1}\a_3)= 2 \cosh\frac{l_3}{2}\cr
&T(\a_3)=2 \cosh \frac{l_{\infty}}{2}\cr
&T(\a_1\b_1\a_1^{-1}\b_1^{-1}\a_2\b_1^{-1}\a_2^{-1}\b_1\a_1)-
 T(\a_1\b_1\a_1^{-1}\b_1^{-1}\a_2\a_1\b_1^{-1}\a_2^{-1}\b_1) =\cr
&\hskip5cm 4 \frac{1}{ \sinh\frac{l_{-\infty}}{2} }\sqrt{s(l_{-\infty},
 l_0,l_1) s(l_{-\infty},l_0,l_2)} \sinh\tau_{-\infty}\cr
&T(\a_1^{-1}\b_1\a_1^{-1}\b_1^{-1}\a_2\b_1\a_1\b_1^{-1})-T(\a_1^{-1}\a_2)=\cr
&\hskip5cm 4 \frac{1}{ \sinh\frac{l_0}{2} }
\sqrt{s(l_{-\infty},l_0,l_1) s(l_{-\infty},l_0,l_2)}\sinh\tau_0\cr
&T(\a_1^{-1}\b_3\a_3\b_3^{-1}\a_2^{-1}\b_1\a_1\b_1^{-1}\a_1^{-1})-
 T(\a_1^{-1}\b_2\a_2^{-1}\b_2^{-1}\a_3) =\cr
&\hskip5cm 4 \frac{1}{ \sinh\frac{l_1}{2} }
\sqrt{s(l_{-\infty},l_0,l_1) s(l_1,l_3,l_\infty)}\sinh\tau_1\cr
&T(\a_1\b_1^{-1}\b_2^{-1}\a_3^{-1}\b_2\a_2\b_1)-
 T(\a_1\b_1^{-1}\a_2\b_2^{-1}\a_3^{-1}\b_2\b_1)=\cr
&\hskip5cm 4 \frac{1}{ \sinh\frac{l_2}{2} }
\sqrt{s(l_{-\infty},l_0,l_2) s(l_2,l_3,l_\infty)}\sinh\tau_2}
$$\hskip1cm
$$
\eqalign{
&T(\a_3\b_3\a_3^{-1}\b_3^{-1}\b_2\a_2^{-1}\b_2^{-1}\a_3)-T(\b_2\a_2^{-1}
\b_2^{-1}\a_3\b_3\a_3^{-1}\b_3^{-1}\a_3)=\cr
&\hskip5cm 4 \frac{1}{ \sinh\frac{l_3}{2} }
\sqrt{s(l_1,l_3,l_\infty) s(l_2,l_3,l_\infty)}\sinh\tau_3\cr
&T(\b_2\a_2^{-1}\b_2^{-1}\a_3\b_3\a_3^{-1}\b_2\a_2\b_2^{-1}
  \a_3^{-1}\b_3^{-1})-
 T(\b_2\a_2^{-1}\b_2^{-1}\a_3\b_3\a_3^{-1}\a_3^{-1}\b_2\a_2
  \b_2^{-1}\b_3^{-1})
=\cr
&\hskip5cm 4 \frac{1}{ \sinh\frac{l_\infty}{2} }
\sqrt{s(l_1,l_3,l_\infty) s(l_2,l_3,l_\infty)}\sinh\tau_\infty.}\eqno(2.6)
$$

\ni The verification of these formulae involves the multiplication of large
numbers of $2\times 2$-matrices, which was done with the help of the algebraic
program Mathematica. One can read off from (2.6) the algebraic form
of the general trace invariant depending on $\tau_j$. If the two pants
meeting at the j'th geodesic cut have boundary components labelled by
$(j,k,l)$ and $(j,m,n)$, say, the corresponding Wilson loop invariant is given
by $4 \sqrt{s(l_j,l_k,l_l)}$ $\sqrt{s(l_j,l_m,l_n)}\sinh\tau_j/\sinh
\frac{l_j}{2}$. No square roots occur in (2.5) because of symmetries
special to the genus-2 case.

One verifies by inspection that these invariants are indeed complete
and global sets of parameters on Teichm\"uller space. The only degeneracies
occur in the singular limit when one or more of the $l_j$ vanish, i.e. part
of the genus-$g$ surface ``pinches off".

The above generalizes straightforwardly to higher genus. The general
expressions for the corresponding $\tau$-dependent Wilson loops in terms of
the generators $\a_i$ and $\b_i$ are given by
\item{a)} for $\tau_{-\infty}$:
$$
T(\a_1\b_1\a_1^{-1}\b_1^{-1}\a_2\b_1^{-1}\a_2^{-1}\b_1\a_1)-
T(\a_1\b_1\a_1^{-1}\b_1^{-1}\a_2\a_1\b_1^{-1}\a_2^{-1}\b_1)
$$
\item{b)} for $\tau_{3i-5}$ ($i=2\dots g-1$):
$$
\eqalign{
&T(\b_{i-1}\a_{i-1}^{-1}\b_{i-1}^{-1}\a_i\b_i\a_i^{-1}\b_i^{-1}\a_{i+1}\b_i
\a_i^{-1}\dots\a_g^{-1}\b_g^{-1})-\cr
&\hskip4cm
T(\b_i\a_i^{-1}\b_i^{-1}\a_{i+1}\b_{i-1}\a_{i-1}^{-1}\b_{i-1}^{-1}\a_i\b_i
\a_i^{-1}\dots\a_g^{-1}\b_g^{-1})}
$$
\item{c)} for $\tau_{3i-4}$ ($i=2\dots g-1$):
$$
T(\b_{i-1}\a_{i-1}\b_{i-1}^{-1}\b_i^{-1}\a_{i+1}^{-1}\b_i\a_i)-
T(\b_{i-1}\a_{i-1}\b_{i-1}^{-1}\a_i\b_i^{-1}\a_{i+1}^{-1}\b_i)
$$
\item{d)} for $\tau_{3i-3}$ ($i=1\dots g-1$):
$$
T(\b_i\a_i^{-1}\b_i^{-1}\a_{i+1}\a_{i+1}\b_{i+1}\a_{i+1}^{-1}\dots\a_g^{-1}
\b_g^{-1})-
T(\a_{i+1}\b_i\a_i^{-1}\b_i^{-1}\a_{i+1}\b_{i+1}\a_{i+1}^{-1}\dots\a_g^{-1}
\b_g^{-1})
$$
\item{e)} for $\tau_\infty$:
$$
\eqalign{
&T(\b_{g-1}\a_{g-1}^{-1}\b_{g-1}^{-1}\a_g\b_g\a_g^{-1}\b_{g-1}
\a_{g-1}\b_{g-1}^{-1}\a_g^{-1}\b_g^{-1})-\cr
&\hskip4cm
T(\b_{g-1}\a_{g-1}^{-1}\b_{g-1}^{-1}\a_g\b_g\a_g^{-1}\a_g^{-1}\b_{g-1}
\a_{g-1}\b_{g-1}^{-1}\b_g^{-1}).}
$$

The dots stand for a product of generators as they occur in the fundamental
relation (1.2). Using this relation, it can easily be verified
that (2.5) and (2.6) are special cases of these expressions.

Given the explicit form for the independent Wilson loops, one may now
translate back and forth between the classical Fenchel-Nielsen and the loop
description. From the point of view of the loop space quantization of 2+1
gravity [4], it is important to know the explicit expression for the natural
volume form on the Teichm\"uller space ${\cal T}_g$, which comes from the
Weil-Petersson symplectic form [11]. Using the self-explanatory notation
$T[l_j]$, $T[\tau_j]$ for the loop invariants given above, the volume form is

$$
\prod_{j=1}^{3g-3}dl_j\,d\tau_j = \frac{1}{\big| \frac{\del (T[l],T[\tau])}
{\del (l,\tau)}\big| } \prod_{j=1}^{3g-3}dT[l_j]\,dT[\tau_j]. \eqno(2.7)
$$

\ni Abbreviating

$$
S[l_i,l_j,l_k]:= T[l_i]^2 +T[l_j]^2 +T[l_k]^2 + T[l_i]\, T[l_j]\, T[l_k] -4
\eqno(2.8)
$$

\ni one finds for the Jacobian of $g=2$

$$
\eqalign{
&\big| \frac{\del (T[l],T[\tau])}{\del (l,\tau)} \big| =
 \sqrt{S[l_{-\infty},l_0,l_\infty]^2 + T[\tau_{-\infty}]^2
(T[l_{-\infty}]^2/4 -1) }\times\cr
&\hskip1cm \sqrt{S[l_{-\infty},l_0,l_\infty]^2 + T[\tau_0]^2
(T[l_0]^2/4 -1) }
 \sqrt{S[l_{-\infty},l_0,l_\infty]^2 + T[\tau_{\infty}]^2
(T[l_{\infty}]^2/4 -1) }, }\eqno(2.9)
$$

\ni for $g=3$

$$
\eqalign{
&\big| \frac{\del (T[l],T[\tau])}{\del (l,\tau)} \big| =\cr
&\hskip1cm \sqrt{S[l_{-\infty},l_0,l_1]S[l_{-\infty},l_0,l_2]
+T[\tau_{-\infty}]^2 (T[l_{-\infty}]^2/4 -1)}\times\cr
&\hskip1cm \sqrt{S[l_{-\infty},l_0,l_1]S[l_{-\infty},l_0,l_2] +T[\tau_0]^2
(T[l_0]^2/4 -1)}\times\cr
&\hskip1cm \sqrt{S[l_{-\infty},l_0,l_1]S[l_1,l_3,l_\infty] +T[\tau_1]^2
(T[l_1]^2/4 -1)}\times\cr
&\hskip1cm \sqrt{S[l_{-\infty},l_0,l_2]S[l_2,l_3,l_\infty] +T[\tau_2]^2
(T[l_2]^2/4 -1)}\times\cr
&\hskip1cm\sqrt{S[l_1,l_3,l_\infty]S[l_2,l_3,l_\infty] +T[\tau_3]^2
(T[l_3]^2/4 -1)}\times\cr
&\hskip1cm\sqrt{S[l_1,l_3,l_\infty]S[l_2,l_3,l_\infty] +T[\tau_\infty]^2
(T[l_\infty]^2/4 -1)}, }\eqno(2.10)
$$

\ni and similarly for higher genus. We observe that the measure, unlike
in its form in terms of Fenchel-Nielsen coordinates, does not factorize
completely.
Given the explicit form of these measures, one may now go ahead and solve a
problem posed in [4], namely that of introducing damping factors to make the
loop transform well-defined. We will not pursue this line of investigation
any further in the present paper.

\vskip1.5cm

\line{\ch 3 Conclusions\hfil}

We have presented above a complete and independent set of (linear
combinations
of) Wilson loop variables for 2+1 gravity on a compact spatial manifold
$\Sigma^g$ of arbitrary genus $g$, and have therefore identified the true
physical degrees of freedom in terms of loop invariants. In the derivation,
crucial use was made of an explicit cross section of the bundle (2.2) over
Teichm\"uller space. The existence of such global loop invariants is possible
because of the topologically trivial nature of the physical configuration
space, and therefore does not generalize to more complicated theories of
connections like for example the $SU(2)$-Yang-Mills theory. Also, our final
solution is simple in that the independent loop variables are not subject to
any inequalities which a priori might have occurred [12].

The elements of the homotopy group $\pi_1 (\Sigma^g)$ going into the
construction of the independent trace invariants are sufficiently complicated
to make it plausible that it would be difficult to obtain similar solutions
by solving the trace identities directly. The algebraic form of our final
solution reiterates the well-known fact that it is usually quite involved to
transform back and forth between the loop and the connection formulation, and
also that much of the original geometric simplicity of the loop formulation
gets lost when solving for the physical degrees of freedom.

\vskip2cm

\line{\ch References\hfil}

\item{[1]} Loll, R.: Chromodynamics and gravity as theories on loop space,
  hep-th/9309056, to appear in Memorial Volume for M.K. Polivanov
\item{[2]} Furmanski, W. and Kolawa, A.: Yang-Mills vacuum: An attempt at
  lattice loop calculus, {\it Nucl. Phys.} B291 (1987) 594-628; Br\"ugmann,
  B.: Method of loops applied to lattice gauge theory, {\it Phys. Rev.}
  D43 (1991) 566-79
\item{[3]} Marolf, D.M.: Loop representations for 2+1 gravity on a torus,
  {\it Class. Quant. Grav.} 10 (1993) 2625-47
\item{[4]} Ashtekar, A. and Loll, R.: New loop representations for 2+1
  gravity, gr-qc/9405031, to appear in {\it Class. Quant. Grav.}
\item{[5]} Witten, E.: 2+1 dimensional gravity as an exactly soluble system,
  {\it Nucl. Phys} B311 (1989) 46-78
\item{[6]} Nelson, J. and Regge, T.: Invariants of 2+1 gravity, {\it Commun.
  Math. Phys.} 155 (1993) 561-8; Quantisation of 2+1 gravity for genus 2,
  Torino preprint DFTT-54-93, gr-qc/9311029
\item{[7]} Okai, T.: An explicit description of the Teichm\"uller space
  as  holonomy representations and its applications, {\it Hiroshima Math.
  J.} 22 (1992) 259-71
\item{[8]} Ashtekar, A., Husain, V., Rovelli, C., Samuel, J., Smolin, L.:
  2+1 gravity as a toy model for the 3+1 theory, {\it Class. Quant. Grav.}
  6 (1989) L185-93; Smolin, L.: Loop representation for quantum gravity in
  2+1 dimensions, in: {\it Knots, topology and quantum field theory},
  Proceedings of the 12th Johns Hopkins Workshop; Ashtekar, A.: Lessons from
  2+1 dimensional quantum gravity, in: {\it Strings 90}, ed. R. Arnowitt et
  al., World Scientific, Singapore
\item{[9]} Goldman, W.M.: Topological components of spaces of representations,
  {\it Invent. Math.} 93 (1988) 557-607
\item{[10]} Abikoff, W.: The real analytic theory of Teichm\"uller
  space,  Lecture Notes in Mathematics, vol. 820, Springer, Berlin, 1980
\item{[11]} Wolpert, S.: On the Weil-Petersson geometry of the moduli
  space of curves, Ann. of Math. 115 (1982) 501-528
\item{[12]} Loll, R.: Loop variable inequalities in gravity and gauge theory,
  {\it Class. Quant. Grav.} 10 (1993) 1471-6

\end